\documentclass{extarticle}
\usepackage{mathpazo}
\usepackage[latin9]{inputenc}
\usepackage[a4paper]{geometry}
\usepackage{color}
\usepackage{amsmath}
\usepackage{amsthm}
\usepackage{amssymb}
\usepackage[numbers]{natbib}
\usepackage[unicode=true,
 bookmarks=false,
 breaklinks=true,pdfborder={0 0 0},pdfborderstyle={},backref=false,colorlinks=true]
 {hyperref}
\usepackage{authblk}
\makeatletter

\theoremstyle{plain}
\newtheorem{lem}{\protect\lemmaname}
\theoremstyle{plain}
\newtheorem{thm}{\protect\theoremname}

\usepackage{amsthm}
\usepackage{latexsym}
\usepackage{amsfonts}
\usepackage{color}
\usepackage{graphicx}
\usepackage{subfigure}
\usepackage{xcolor}
\hypersetup{
	colorlinks,
	linkcolor={red!50!black},
	citecolor={blue!50!black},
	urlcolor={blue!80!black},
               }

\makeatother

\providecommand{\lemmaname}{Lemma}
\providecommand{\theoremname}{Theorem}

\title{Comment on `Product states and Schmidt rank of mutually unbiased
bases in dimension six'}
\author[a]{Daniel McNulty\thanks{daniel.mcnulty@uniba.it}}
\author[b]{Stefan Weigert \thanks{stefan.weigert@york.ac.uk}}

\affil[a]{Dipartimento di Fisica, Universit\`{a} di Bari, Bari, Italy}
\affil[b]{Department of Mathematics, University of York, York, UK}
\begin{document}
\date{April 17, 2025}
\maketitle

\begin{abstract}
A lemma by Chen \emph{et al.} \href{https://doi.org/10.1088/1751-8121/aa8f9e}{[\emph{J. Phys. A: Math. Theor.} {\bf 50}, 475304 (2017)]}
provides a necessary condition on the structure
of any complex Hadamard matrix in a set of four mutually unbiased
bases in $\mathbb{C}^{6}$. The proof of the lemma is shown to contain
a mistake, ultimately invalidating three theorems derived in later
publications.
\end{abstract}
\vspace{0.7cm}
It is unknown whether complete sets of $(d+1)$ mutually unbiased
(MU) bases exist in a complex Hilbert space $\mathbb{C}^{d}$ if the
dimension $d\in\mathbb{N}$ does \emph{not} equal the power of a prime.
The simplest case of this long-standing open problem arises when $d=6$,
the smallest composite dimension that is not a prime-power. It has
been conjectured that no more than three MU bases exist for $d=6$.
In the space $\mathbb{C}^{6}$, any orthonormal basis MU to the standard
basis corresponds to a complex Hadamard matrix of order six. One
strategy to prove the conjecture is to derive an exhaustive\footnote{A (not necessarily exhaustive) list of the known complex Hadamard matrices of order six can be found in the online catalogue \citep{catalog}. It also contains the definitions of the Hadamard matrices appearing throughout this note, such as $D_{6}^{(1)}$, $B_{6}^{(1)}$, and $M_{6}^{(1)}$.} list of $6\times6$ Hadamard matrices and show that none of them
can be part of a \emph{quadruple} of MU bases.

Along these lines, Chen \emph{et al.} \citep{chen17} have ruled out
the existence of quadruples of MU bases that contain Hadamard matrices
of a specific form. Unfortunately, one part of a lemma they present
is marred by an erroneous proof.
\begin{lem}[\citep{chen17}, Lemma 11(v) Part 6]
\label{lem} If a set of four MU bases in dimension six exists, none
of the Hadamard matrices from the set contains a real $3\times2$
submatrix.
\end{lem}
In its original formulation, the lemma considers a ``MUB trio'', which
is a set of three mutually unbiased Hadamard matrices. Equivalently,
we consider a set of four MU bases and assume that one basis is the
standard one.

The proposed proof proceeds by contradiction: a set of four MU bases is assumed
to exist that contains both the identity matrix $\mathbb{I}$ and
a complex Hadamard matrix $H$ with a real submatrix of size $3\times2$.
By applying suitable row and column permutations to the four MU bases,
as well as rephasing individual basis vectors, one can ensure that
the first row and column of $H$ have entries all equal to $1/\sqrt{6}$
and that its upper left $3\times2$ matrix is real, i.e. 
\begin{equation}
\frac{1}{\sqrt{6}}\left(\begin{array}{cc}
1 & 1\\
1 & y\\
1 & x
\end{array}\right),\qquad y,x\in\{\pm1\}\,.\label{eq:submatrix}
\end{equation}

Let $y=1$. If the submatrix (\ref{eq:submatrix}) has rank one, which
occurs for $x=1$, the matrix $H$ cannot be a member of a MU quadruple,
as shown in Part 2 of Lemma 11(v) of Ref. \citep{chen17}. In this
case, the matrix $H$ must contain a $3\times3$ unitary matrix. If
it does, the pair $\{\mathbb{I},H\}$ would be equivalent to a pair
of MU product bases which, however, cannot be extended to four MU
bases as shown in \citep{mcnulty12a,mcnulty12b}. Therefore, one must
have $x=-1$. If $y=-1$, both choices $x=\pm1$ also lead, upon suitably
permuting rows and rephasing the second vector, to the case $(y,x)=(1,-1)$.

Now, given $(y,x)=(1,-1)$, orthogonality of the first two columns
of $H$ implies that the last three elements of the second column
must be given by $(-1,s,-s)/\sqrt{6}$, where $s$ is a complex number
of modulus one. The third column vector (labelled $v$) of $H$ must
be orthogonal to both the first and the second column of $H$. According
to Ref. \citep{chen17}, these conditions on the vector $v$ imply
that its ``third and sixth elements must be zero'' \citep[p. 24]{chen17}.
Since $H$ had been assumed to be a Hadamard matrix, a contradiction
is reached that is sufficient to complete the proof of Lemma \ref{lem}.
However, we are unable to confirm that two components of the vector
$v$ must vanish. Therefore, it is not possible to rule out the existence
of MU quadruples containing Hadamard matrices with real $3\times2$
submatrices.

An example shows explicitly that the final step of the argument cannot
be correct. Consider the one-parameter family of \emph{symmetric}
Hadamard matrices
\begin{equation}
M_{6}^{(1)}\equiv M_{6}(a)=\frac{1}{\sqrt{6}}\left(\begin{array}{cccccc}
1 & 1 & 1 & 1 & 1 & 1\\
1 & -1 & a & a & -a & -a\\
1 & a & b & c & d & e\\
1 & a & c & b & e & d\\
1 & -a & d & e & f & g\\
1 & -a & e & d & g & f
\end{array}\right)\,,
\end{equation}
where $a=e^{it}$ and $t\in(\pi/2,\pi]\cup(3\pi/2,2\pi]$. The entries
$b,c,d,e,f$ and $g$ are functions of $a$ defined in Ref. \citep{catalog}.
Multiplying the second column of this matrix by the complex conjugate
of $a$ produces a real $4\times2$ submatrix in the lower left corner.
Suitable row permutations turn the upper left $3\times2$ matrix into
the expression in \eqref{eq:submatrix} with $(y,x)=(1,-1)$. Multiplying
the remaining columns by suitable phase factors, we arrive at a matrix
$H$ that, according to the analysis given in the previous paragraph,
cannot exist. In addition, there is no proof that excludes the matrices
$M_{6}(a)$ from appearing in a set of four MU bases (although numerical
evidence suggests that for some values of the parameter $a$, pairs
of the form $\left\{ \mathbb{I},M_{6}(a)\right\} $ cannot even be
extended to a triple of MU bases \citep{goyeneche2013}). Note that
$M_{6}(a)$ cannot, in general, contain three columns that form product
vectors, including after row and column permutations, therefore Part
4 of Lemma 11(v) of Ref. \citep{chen17} does not apply.

We are aware of at least three theorems on the existence of MU quadruples
that build on the now unverified lemma, derived in Refs. \citep{liang21,liang19,chen21}.
Thus, their validity is called into question. As before, we assume
that any set of four MU bases we consider contains the standard basis.
\begin{thm}[\citep{liang21}, Theorem 10]
\label{thm1} If a set of four MU bases in dimension six exists,
none of the Hadamard matrices from the set contains more than 22 real
entries.
\end{thm}
The proof of this theorem explicitly uses Lemma \ref{lem} to exclude
Hadamard matrices with more than 22 real entries from appearing in
quadruples of MU bases.\footnote{Some of the authors of \citep{liang21} have privately communicated
an alternative proof of Thm. \ref{thm1} that is independent of Lemma
\ref{lem} and awaits publication.}

A restriction on the type of $H_{2}$-reducible Hadamard matrices
that are permitted in MU quadruples was derived in Ref. \citep{liang19}.
A complex Hadamard matrix of order six is $H_{2}$\emph{-reducible}
if it can be partitioned into nine $2\times2$ blocks, each proportional
to a Hadamard matrix of order two. The complete set of $H_{2}$-reducible
matrices is known as the three-parameter \emph{Karlsson }family (cf.
\citep{catalog}).
\begin{thm}[\citep{liang19}, Theorem 12 \& Lemma 13]
\label{thm2} A $H_{2}$-reducible matrix in a set of four MU bases
contains exactly nine or eighteen $2\times2$ submatrices proportional
to Hadamard matrices.
\end{thm}
This claim relies directly on Lemma \ref{lem}: the non-existence
of specific submatrices of size $3\times2$ is used to limit the number
of $2\times2$ Hadamard submatrices.\footnote{The proof of Thm. \ref{thm2} also contains an inconsistency unrelated
to Lemma \ref{lem}. Consider, for example, the one-parameter family
of self-adjoint Hadamard matrices $B_{6}^{(1)}$. Members of this
family contain no $3\times2$ real submatrix but more than eighteen
$2\times2$ Hadamard submatrices, as can be seen by inspection. However,
by the argument applied to prove Thm. \ref{thm2} (which relies only
on Lemma \ref{lem} to restrict the bases), $B_{6}^{(1)}$ should
not be excluded from appearing in an MU quadruple. It is therefore
feasible---regardless of the veracity of Lemma \ref{lem}---that
an MU quadruple contains a Hadamard matrix with more than eighteen
$2\times2$ Hadamard submatrices.}

Finally, a theorem in Ref. \citep{chen21} further limits the number
of $2\times2$ Hadamard submatrices of any $H_{2}$-reducible matrix
in a quadruple of MU bases. This property is then used to severely
restrict the types of Hadamard matrices that may figure in such a
set.
\begin{thm}[\citep{chen21}, Theorems 7, 8 \& 9]
\label{thm3} A $H_{2}$-reducible matrix in a set of four MU bases
contains exactly nine $2\times2$ submatrices each proportional to
a Hadamard matrix. Thus, members of the families $D_{6}^{(1)}$, $B_{6}^{(1)}$,
$M_{6}^{(1)}$ and $X_{6}^{(2)}$ do not figure in a quadruple of
MU bases.
\end{thm}
Thm. \ref{thm2} and Lemma \ref{lem} are required to prove Thm. \ref{thm3}.
Hence, the restrictions claimed in Thm. \ref{thm3} cannot be upheld.

Without a proof of Thm. \ref{thm3}, only a few constraints on the
types of complex Hadamard matrices that may figure in an MU quadruple
remain known. The isolated matrix $S_{6}$ does not extend to a triple,
let alone a quadruple. Quadruples do not contain members of the two-parameter
Fourier family $F_{6}^{(2)}$, as shown rigorously by a combination
of analytic estimates and numerical evidence \citep{jaming10} as
well as a proof using Delsarte's bound \citep{matolcsi15}. To the
best of our knowledge \citep{mcnulty+24}, no non-existence proofs
for other families are known.

\subparagraph*{Acknowledgements}

We would like to thank Lin Chen and Li Yu for comments on a draft
of this note. D.M. acknowledges support from PNRR MUR Project No. PE0000023-NQSTI.

\end{document}